\documentclass[a4paper, 11pt]{article}
\usepackage{graphicx}
\usepackage{epstopdf}
\usepackage{cite}

\usepackage{bm}
\usepackage[english]{babel}
\begin{document}

\title{A non equilibrium Ising model of turbulence}

\author{Davide Faranda\\
              LSCE-IPSL, CEA Saclay l'Orme des Merisiers,\\ CNRS UMR 8212 CEA-CNRS-UVSQ,\\ Universit\'e Paris-Saclay, 91191 Gif-sur-Yvette, France\\
              Tel.: +33-169081142\\
              Fax:  +33-169087716\\
              davide.faranda@lsce.ipsl.fr\\
          	   Martin Mihelich \\
   		      Laboratoire SPHYNX, SPEC,   CEA Saclay, CNRS UMR 3680,\\ Universit\'e Paris-Saclay, 91191 Gif-sur-Yvette, France\\
           B{\'e}reng{\`e}re Dubrulle \\
   		      Laboratoire SPHYNX, SPEC,   CEA Saclay, CNRS UMR 3680,\\ Universit\'e Paris-Saclay, 91191 Gif-sur-Yvette, France
}

\maketitle

\begin{abstract}

We introduce a  model of interacting lattices at different resolutions  driven by the two-dimensional Ising dynamics with a nearest-neighbor interaction. We study this model both with tools borrowed from equilibrium statistical mechanics as well as non-equilibrium thermodynamics. Our findings show that this model keeps the signature of the equilibrium phase transition. Moreover  the critical temperature of the equilibrium models correspond to the state maximizing the entropy and delimits two out-of-equilibrium regimes, one satisfying the Onsager relations for systems close to equilibrium and one resembling convective turbulent states. Since the model preserves the entropy and energy fluxes in the scale space, it seems a good candidate for parametric studies of out-of-equilibrium turbulent systems.
\end{abstract}

\section{Introduction}

Equilibrium systems can often be described through well founded statistical mechanics, built up from the classical laws of thermodynamics. In this sense, they represent an exception with respect to most systems found in nature, which are subject to flux of matter and energy to and from other systems  and/or to chemical reactions. Most systems are therefore intrinsically out-of-equilibrium and their description often remains beyond the scope of present statistical mechanics~\cite{lebon2008understanding}. Among all out-of-equilibrium systems, we may single out those leading to (out-of-equilibrium) steady states, called NESS, which exhibit a constant average energy. Intense research efforts have thus focused on understanding how much such states differ from the equilibrium states of corresponding ideal systems (without say, forcing and dissipation). For example, there is now strong evidence that the large scales of the steady states of 2D or quasi-2D turbulent flows can be described as equilibrium states of the corresponding Euler equation~\cite{bouchet2012statistical,thalabardstatistical}, whereas such flows are subject to non-zero energy or enstrophy fluxes through scales, characteristic of out-of-equilibrium dynamics. On the other hand, there has been recent numerical and theoretical evidence of steady states in out-of-equilibrium systems that do not correspond to any equilibrium states, such as an Ising model driven by a temperature gradient~\cite{pleimling2010convection}, and a quantum open system in contact with two reservoirs~\cite{huang2011entropy}. One way to get some insights about why and how NESS approach their equilibrium counterpart relies on the study of simple toy models which are analytically tractable. ASEP (symmetric simple exclusion processes, also named zero range processes) have become a paradigm of non-equilibrium toy models on a lattice~\cite{derrida2002large} which mimic diffusion phenomena. Such models are however not appropriate to understand the turbulent NESS, because it lacks an essential ingredient: the development of an energy cascade over a wide range of scales. This was recognized by \cite{bertin2009far}, who devised a dissipative forced zero range process on a Cayley (self-similar) tree identified in the article to a scale space. This process exhibits a transition between a quasi-equilibrium regime and a far-from equilibrium regime where net fluxes through scales were observed. However, this example does not exactly describe turbulent NESS, for which both the non-zero energy flux and the equilibrium state are superposed.

In the present article, we study a new toy model of ``turbulent process'' based on the Ising model~\cite{ising1925beitrag,mccoy1973two}. It  approximates magnetic dipole moments of atomic spins through discrete variables with Boolean-like distribution (+1 or −1), interacting with their nearest neighbors on a lattice. The properties of the model at thermal equilibrium are well established and allow for the identification of second order phase transitions in 2D~\cite{onsager1944crystal,onsager1949nuovo}. The idea of exploring the existence of non-equilibrium steady states (NESS) of the Ising model as a paradigm for other phenomena hails from the work of  \cite{glauber1963coherent}, who devised simple models for out-of-equilibrium dynamics when in contact with a thermostat. Other modifications of the Ising model in the same spirit have been suggested by~\cite{garrido1987ising,garrido1990long,cheng1991long,praestgaard2000lattice,pleimling2010convection,maes1991anisotropic,pinto2014critical} where the addition of a temperature gradient within the lattice sets the system out of equilibrium. However, none of these variants of the Ising model so far contain the dynamics of energy or enstrophy fluxes through scales. In order to take into account this feature, we propose a model of two interacting lattices. Lattice A is the reference: its dynamics is iterated at temperature $T_A$, then the configuration is copied to lattice B and an iteration is performed at $T_B$. Finally the first lattice is updated with the changes applied to lattice B and currents are computed as the number of spins changed within these dynamical steps. Coarse graining effects are monitored during this procedure by downgrading the resolution of lattice B.\\

The paper is organized as follows: we introduce the two-lattice model in Section~\ref{twolat}. We then perform two different analyses. Section~\ref{equianalys} is devoted to the study of possible traces of classical equilibrium phase transitions in the model dynamics. We then use non-equilibrium tools in Section~\ref{nonequianalys} to study the thermodynamics of the system. We will show that coarse grain effects alone introduce non equilibrium dynamics on the model and that such dynamics depend both on the temperature gradient between the two lattices A and B but also on the temperature $T_A$ of the fine grained grid. Finally, we discuss in Section~\ref{disc} the possible physical implications of our findings.\\

\section{Model of two interacting lattices at different temperatures.}
\label{twolat}

\begin{figure}
\includegraphics[width=0.65\textwidth]{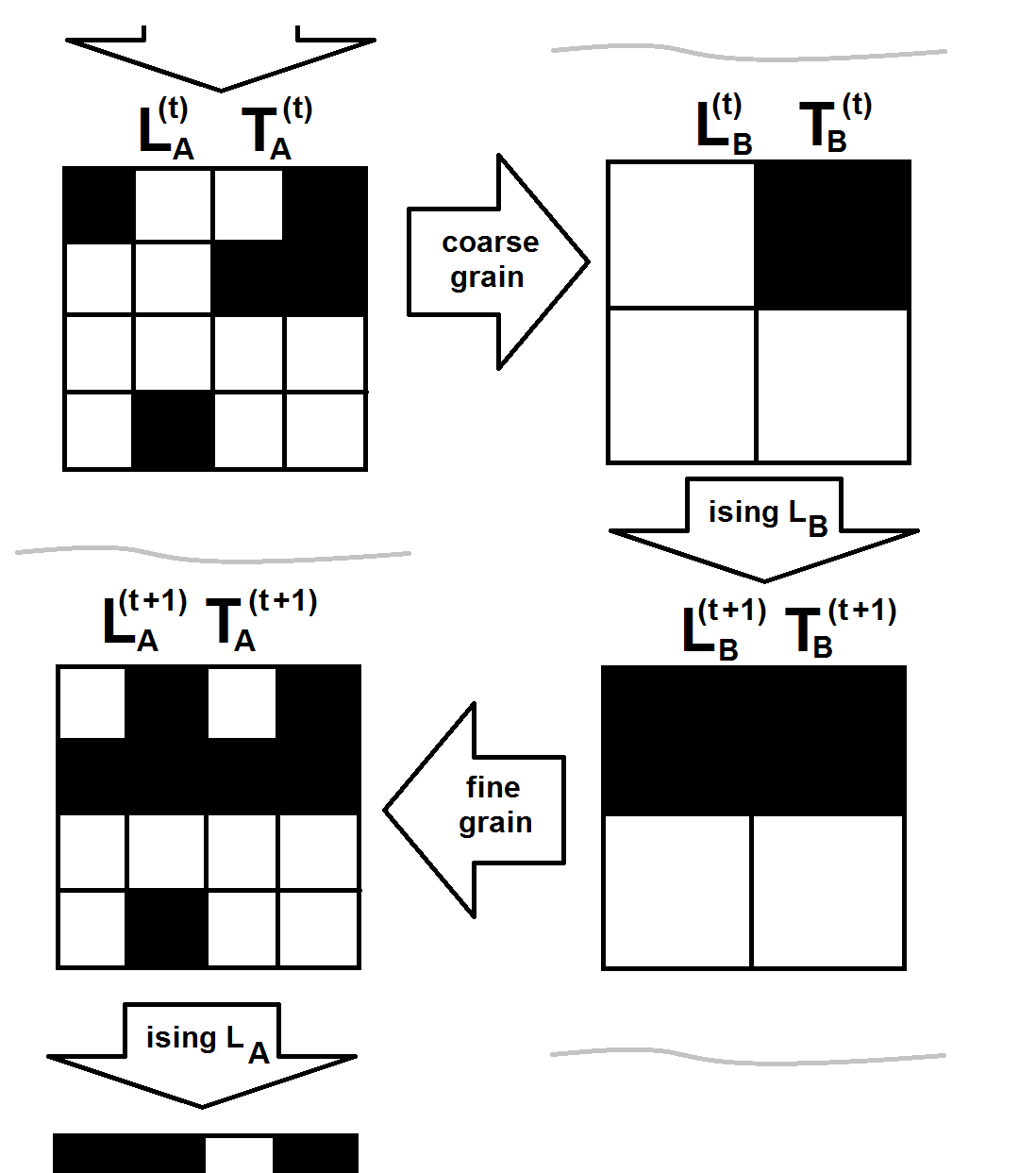}
\caption{Schematic representation of the two lattices model. In this representation the size $L_B$ of grid $B$ is smaller than $L_A$. Vertical steps are dynamical and related to the iteration of the Ising dynamics at temperature $T_A$ or $T_B$. The horizontal steps describe the way the operations of coarse grain and fine grain are performed.} 
\label{schema}
\end{figure}

We introduce a model consisting of two square lattices $A$ and $B$, each of them individually obeying the classical 2-D Ising Kawasaki (conserving average magnetization) dynamics with a nearest-neighbor interaction. The two lattices evolve at different temperatures $T_A$ and $T_B$ and may be of different sizes $L_A$ and $L_B$. A schematic representation of the dynamics of the model is reported in Figure~\ref{schema}. First, lattice $A$ is updated with a Monte Carlo step at temperature $T_A$~; then, if $L_A\neq L_B$, the coarse grain operation is performed: for each spin of the lattice at lower resolution we count the number of corresponding spins in at full resolution and, if the majority is positive (black in the figure) we assign a positive spin to the corresponding coarse grain position. We then perform a Monte Carlo step at the temperature $T_B$ on lattice $B$. Eventually we update grid $A$ by reversing all the spins which correspond to a change in lattice $B$. 

Non-equilibrium effects can be tracked measuring non-zero average fluxes $j(t)$ circulating in the two-lattice system. The most natural definition for $j$ computes the number of spins changed on grid $A$ as an effect of the coarse grain and Monte Carlo steps performed on grid $B$ at temperature $T_B$. In the following we will describe some of the system properties, the role of non-equilibrium effects and of the phase transitions. We will see that there is no trivial complete description of the non-equilibrium effects using solely statistical tools devised for equilibrium dynamics. However, relevant quantities of equilibrium dynamics still play an important role when the system is set out of equilibrium.

\section{Equilibrium analysis and phase transitions}
\label{equianalys}

\begin{figure}
\includegraphics[width=0.45\textwidth]{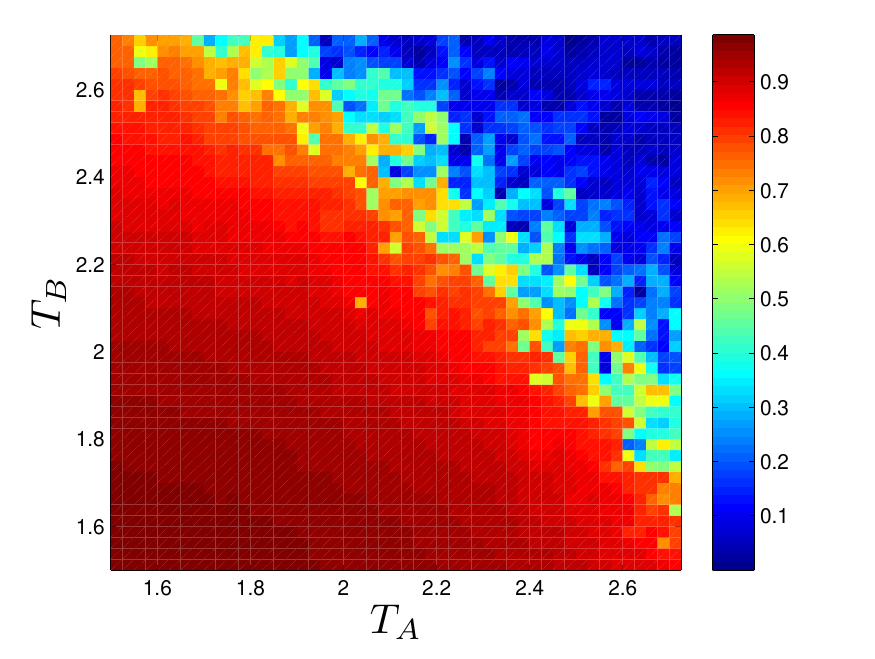}\includegraphics[width=0.45\textwidth]{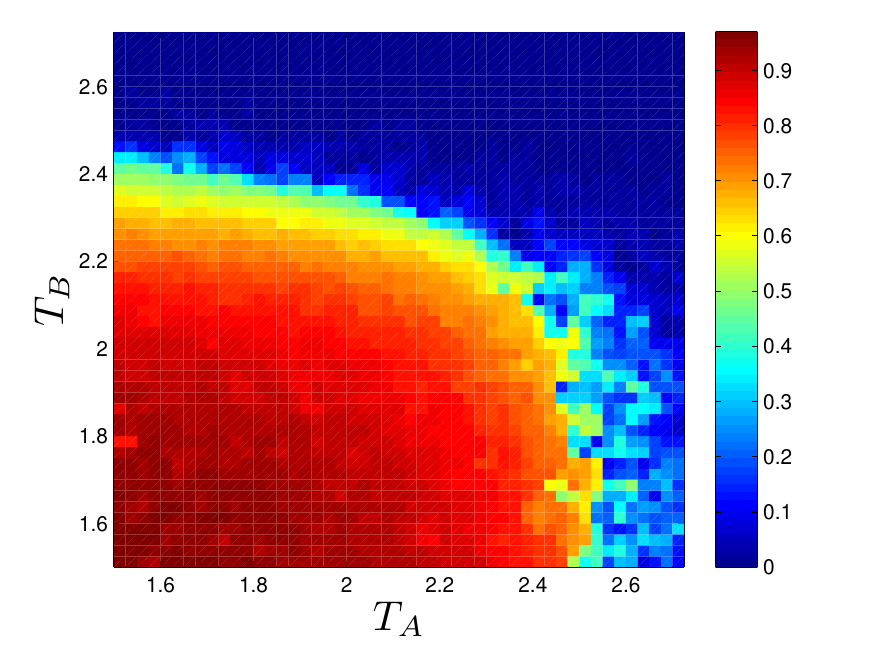}\\
\includegraphics[width=0.45\textwidth]{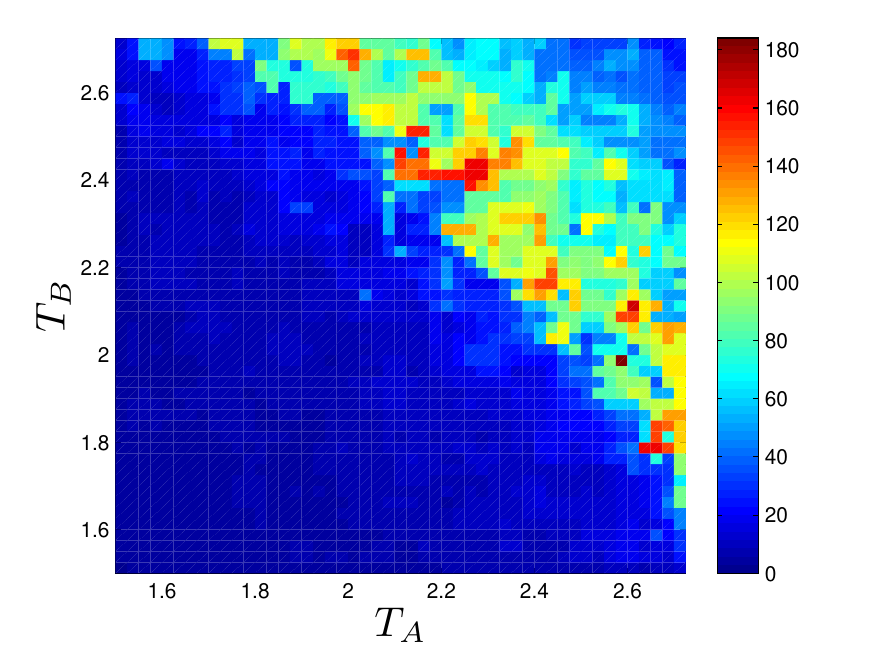}\includegraphics[width=0.45\textwidth]{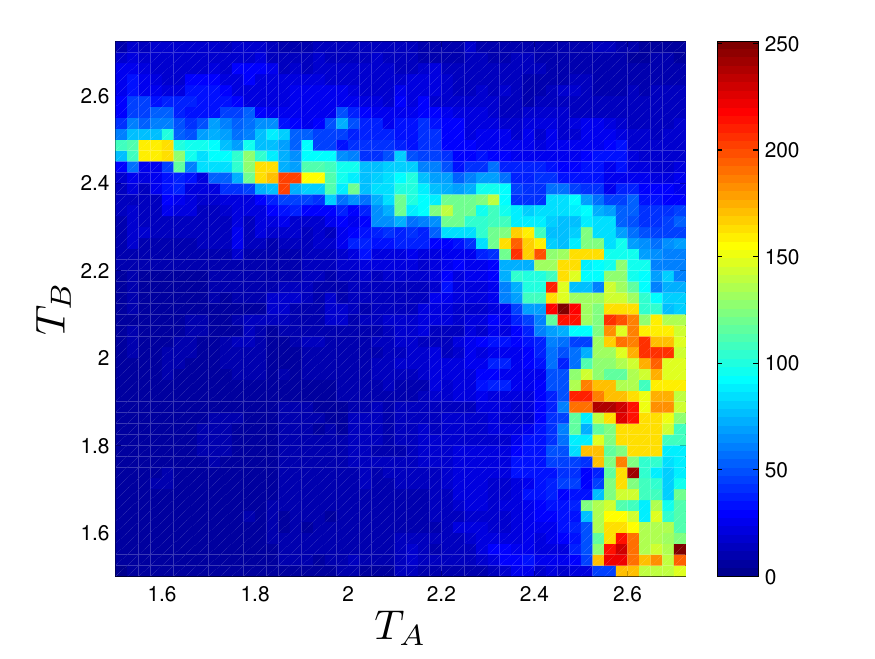}
\caption{ Average magnetization $ \langle \bar{M} \rangle $  (upper panels) and first zero of the autocorrelation function $C(\tau)$ as a function of the temperatures $T_A$ and $T_B$.  Left panel: lattice without coarse grain $L_A=L_B=64$. Right panel: coarse grain with $L_A=64, L_B=8$} 
\label{dyna}
\end{figure}

We have investigated the dynamics of the systems for different temperatures $T_A$ and $T_B$ and different sizes $L_A$ and $L_B$ from $L=8$ to $L=256$. On this range we have found that results do not qualitatively depend on the resolution. Therefore, in the following, we will consider two typical experiments: one where the two grids are at the same resolution but at two different temperatures $L_A=L_B=64$ and  another one where the two grids are at different temperatures and resolutions $L_A=64$ and $L_B=8$.  For each experiment we compute the relevant quantities of the Ising dynamics on 1000 steps as described in Fig.~\ref{schema}. The temperature range analyzed is $(T_A,T_B) \in [T_c-1, T_c+1]^2$, $T_c = 2 / \ln(1 + \sqrt{2})$ being the reduced critical temperature for two-dimensional square lattices. \\
We first notice that the system sets very quickly in a stationary state where the magnetization and the energy fluctuate around a well-defined value. Hence, it is interesting to see whether such steady states still undergo a phase transition at a certain temperature set $(T_A,T_B)$ ; and to assess how such a transition depends on the coarse grain effects. In order to do so, for each experiment we have computed the time and space average magnetization $ \langle \bar{M} \rangle $, and the first zero of the auto-correlation function $C(\tau)$ of the time series of magnetization. For a classical Onsager square lattice, $C(\tau)$ is known to decrease exponentially with a characteristic time which diverges following a power law near the critical temperature $T_c$, whereas $ \langle \bar{M} \rangle $ satisfies a critical power-law scaling near the same critical temperature $T_c$. Both quantities are represented in Fig.~\ref{dyna} for the non coarse grained experiment (left plots) and the coarse grained experiments (right).\\

\begin{figure}
\includegraphics[width=1\textwidth]{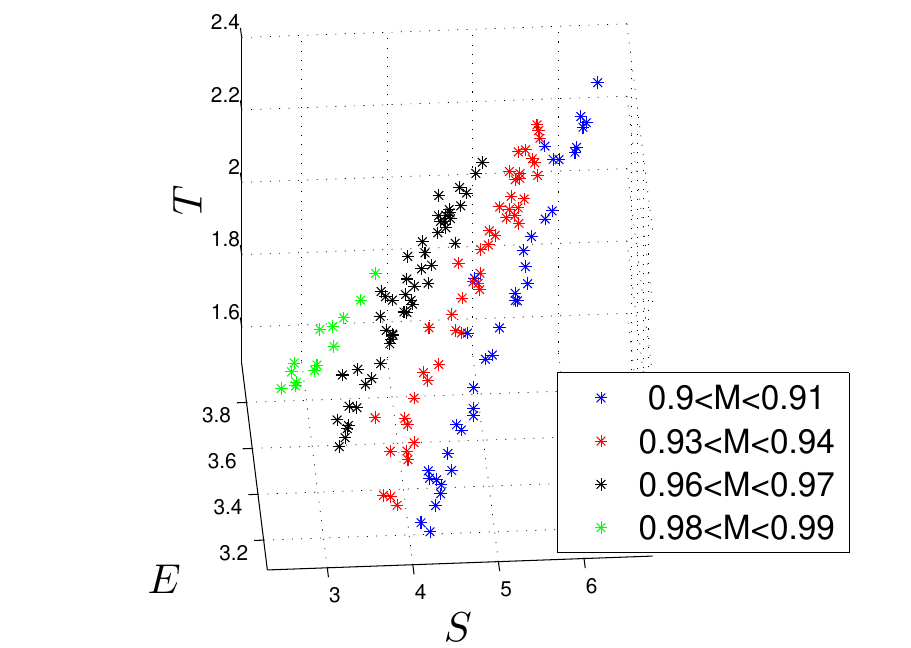}
\caption{Numerical verification of the Onsager relations close to equilibrium. The temperature $T$ is a good  linear function of the Shannon Entropy $S$ and the energy $E$, when the magnetization $M$ is kept constant. Different colors correspond to different $M$ intervals as described in the legend. }
\label{EnergyEntropy}
\end{figure}

We first observe that the absolute value of $ \langle \bar{M} \rangle $ (upper panels of Fig. \ref{dyna}) reproduces the equilibrium Ising critical transition but with a higher pseudo-critical temperature $T_c(T_A,T_B)$.  When no coarse grain is applied to the grid $L_B$ the picture looks  symmetric with respect $T_A$ and $T_B$.  In this sense, one could define  $T_c(T_A,T_B)$ by following the curve  on the $T_A, T_B$ plane where $ \langle \bar{M} \rangle $ changes abruptly. On the other hand, when the resolution of $L_B$ is decreased (top right panel of the same figure), this curve does not look symmetric with respect to $T_A$ and $T_B$. The transition looks sharper when the temperature of the coarse grained lattice $L_B$ is varied. Our model also respects the symmetry of the magnetization with respect to zero: positive and negative magnetization are obtained with equal probability. \\
Further signatures of phase transitions can be found computing the auto-correlation function $C(\tau)$ of the magnetization time series $\langle M(t) \rangle$. The first zero of $C(\tau)$ is reported in the lower panels of Fig.~\ref{dyna}. In both cases we observe an increase of the resulting time scale in a region of the $(T_A, T_B)$ plane which can be readily identified as a critical temperature. This region matches the region of maximal variations of the average magnetization $ \langle \bar{M} \rangle $.\\

We now turn to analyze the effect of the temperature as a function of the entropy $S$ and the energy  $E$. In fact, as predicted by Onsager, if the system remains close to equilibrium, one must find: $\partial E/\partial S|_M=T_A$, where $M$ is the magnetization at the end of the simulation. The entropy used here is the usual Shannon Entropy computed in information theory which measures the \textit{disorder} of lattice $A$ at the end of the simulation:
\[ 
	S=  -\sum_{i,j}p(i,j)log_2(p(i,j)) 
\]
with $p(i,j)$ the probability of observing a spin up or down at the position $(i,j)$. The linearity of such a relation can be verified in Fig.~\ref{EnergyEntropy} where the quantities $E,S,T_A$ are plotted for four different values of $M$.\\

In the next section we will perform a  non equilibrium analysis considering thermodynamic quantities such as the entropy and the currents, and understand whether such a phase transition is still relevant when the system is set out of equilibrium.

\section{Non-equilibrium analysis}
\label{nonequianalys}

\begin{figure}
\includegraphics[width=0.45\textwidth]{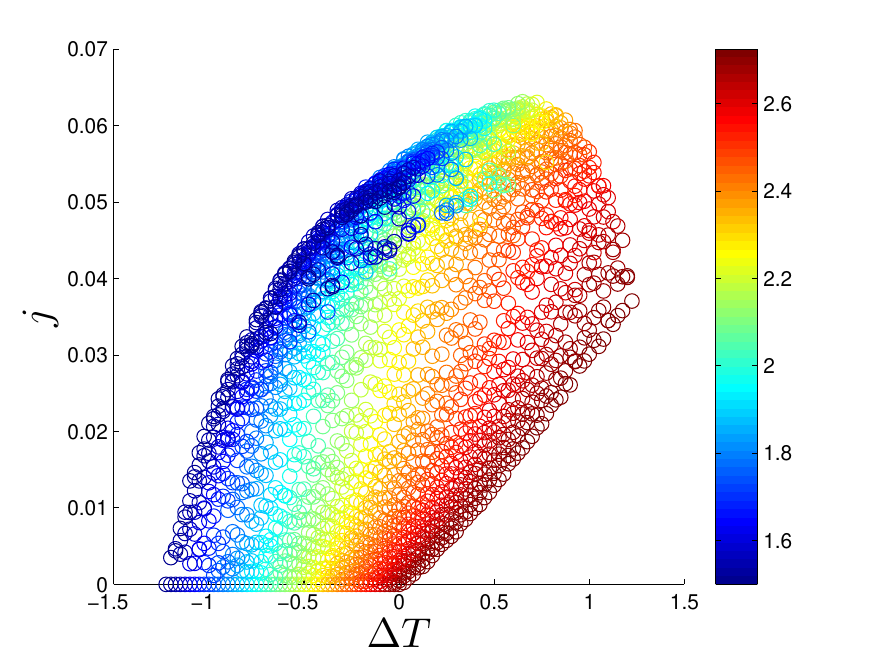}\includegraphics[width=0.45\textwidth]{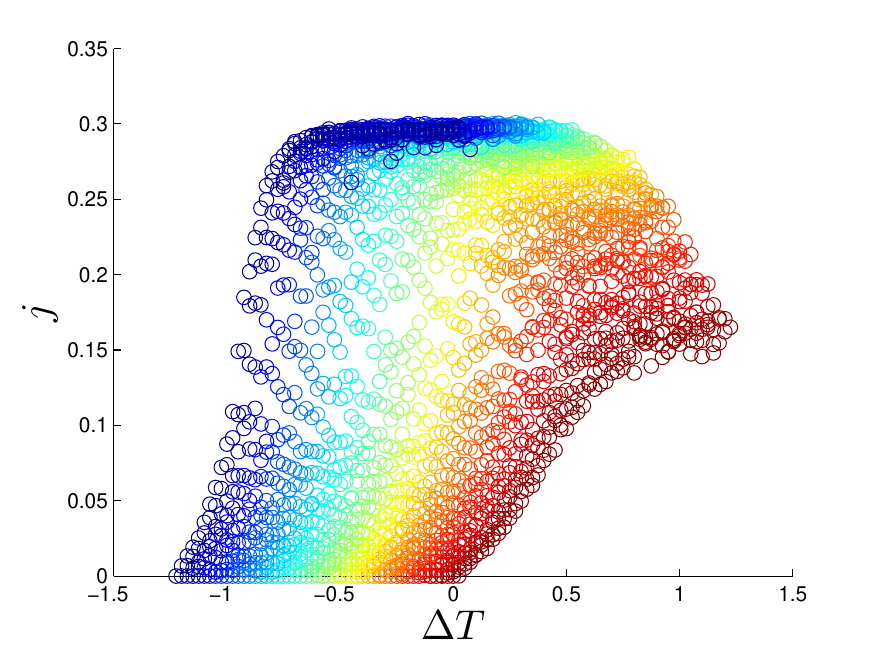}
\caption{Current $j$ between grids $B$ and $A$, as a function of the temperature gradient $\Delta T=T_B-T_A$. The color scale represents temperature $T_B$. Left: lattices without coarse grain $L_A=L_B=64$. Right: coarse grain with $L_A=64, L_B=8$} 
\label{Currents}
\end{figure}

We start the analysis verifying that the model proposed in this section possesses a genuine non equilibrium dynamics and then comment on some of its relevant thermodynamic properties. We will use the definition of currents already described above, i.e. we define $j$ as the number of spins  changed on lattice $A$ after the steps performed on lattice $B$. We rescale $j$ to include only the effects of the temperature change.\\
In Fig.~\ref{Currents} we present the behavior of $j$ for the two experiments detailed before: no coarse grain on the left panels and coarse grain on the right. Colors represent the temperature $T_B$.  For the non coarse grained dynamics we immediately see that, for each temperature --~each color~-- $T_B$ the current $j$ increases proportionally to $\Delta T = T_B - T_A$. This is the signature of Fourier's relation when the system is driven out of equilibrium and the \textit{heat transfer} can be considered conductive. However, for higher temperature gradients and depending on $T_B$ this regime is broken and currents saturate at a specific value. This phenomenon is facilitated for coarse grained dynamics (right panel) where the saturation is indeed observed at all temperatures $T_B$. \\

\begin{figure}
\includegraphics[width=0.5\textwidth]{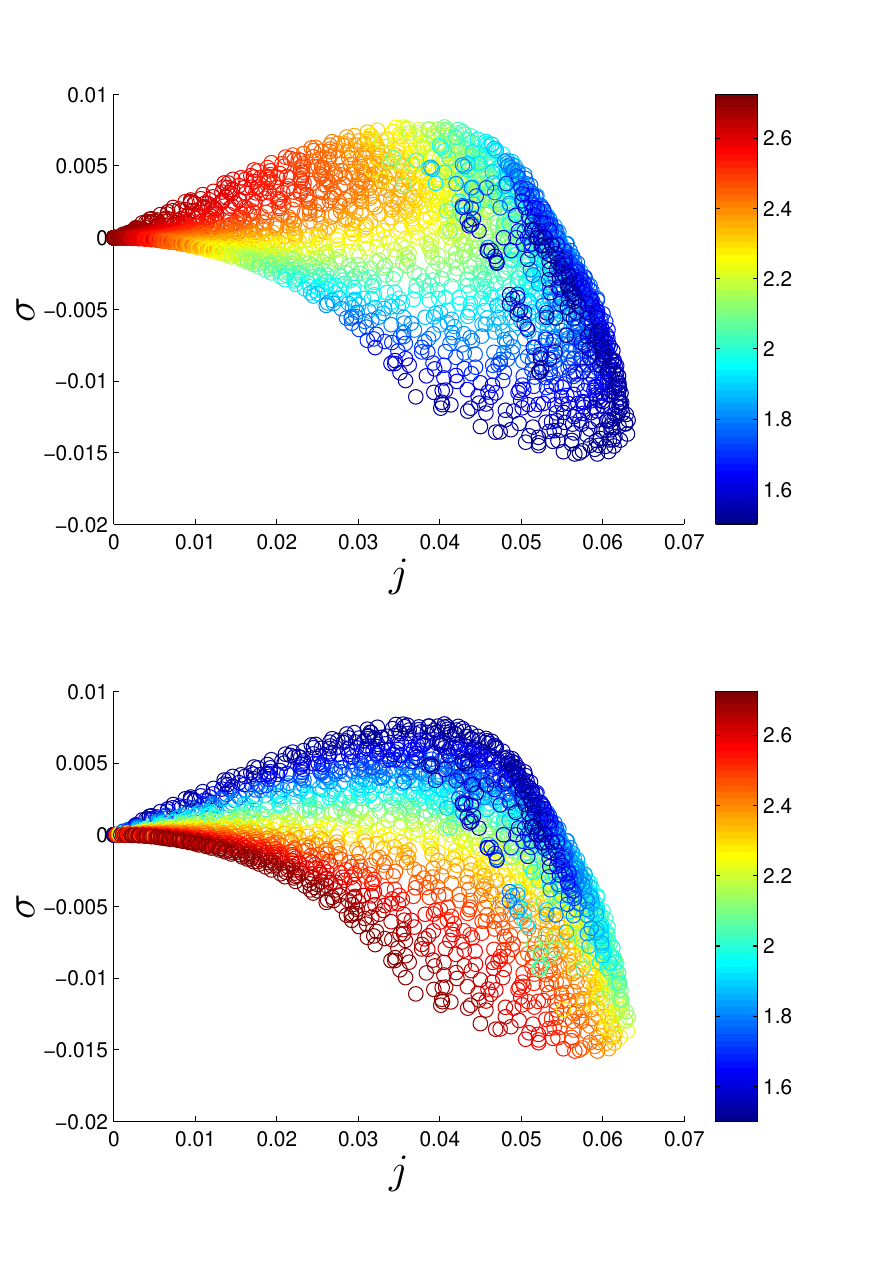}\includegraphics[width=0.5\textwidth]{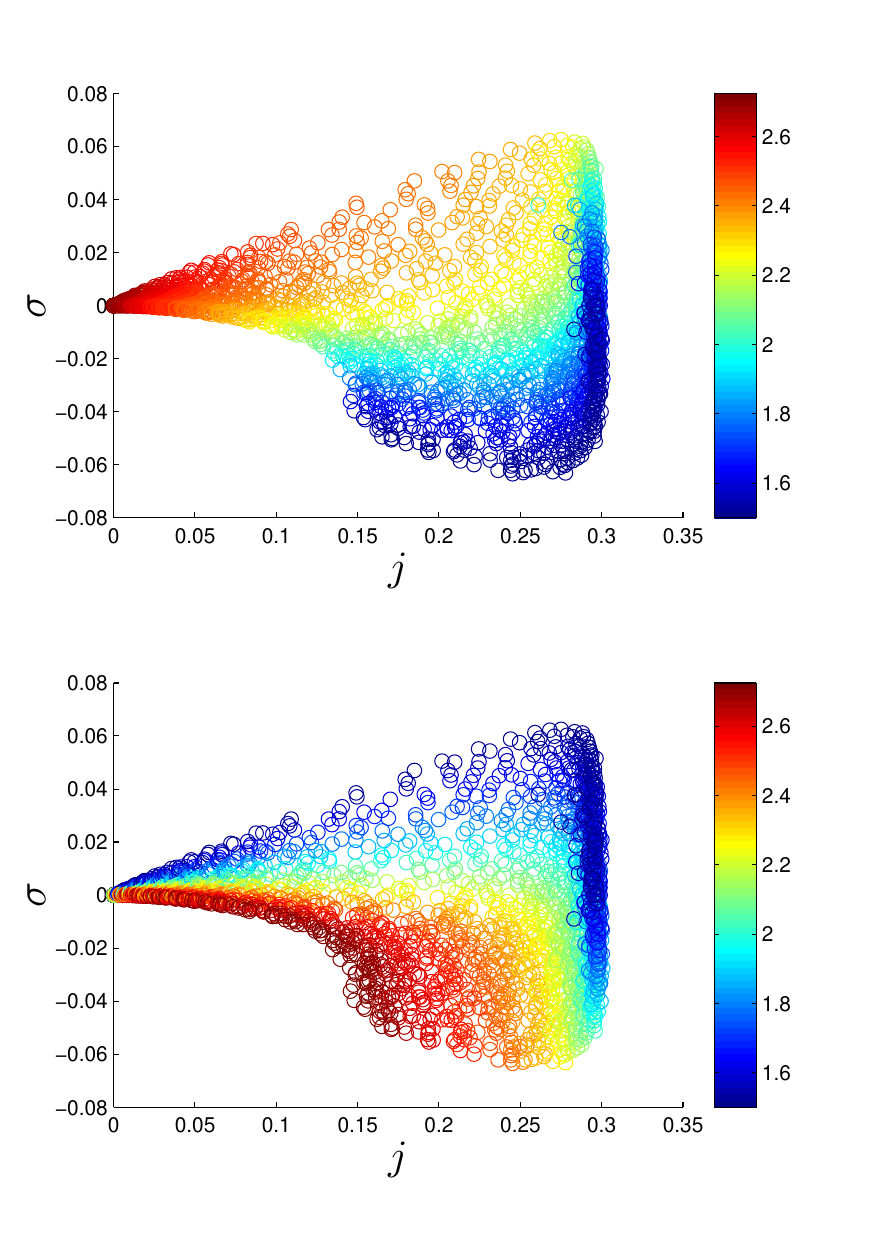}
\caption{Entropy production $\sigma$ as a function of the currents $j$. Left panels: lattice without coarse grain $L_A=L_B=64$. Right panels: coarse grain with $L_A=64, L_B=8$.  The color scale represents $T_A$ (upper panels) and $T_B$ (lower panels).} 
\label{Entropy}
\end{figure}

Another important proof of far-from-equilibrium properties relies on the existence of a maximum of  entropy production and the relation linking this maximum to the temperatures $T_A$ and $T_B$ and to the currents and their conductive/convective nature. In fact, as pointed out in~\cite{mihelich2014maximum,mihelich2014towards},  for other out of equilibrium models such as the zero range process, a maximum of entropy production is clearly recognizable and falls between these two regimes. In our model we will define the macroscopic Entropy Production $\sigma$ as the product of the fluxes by the corresponding thermodynamics forces responsible for such fluxes. It is important to point out that this entropy is different from the microscopic entropy defined above. In terms of currents and temperatures the Entropy Production is written as:

\begin{equation}
\sigma=j\left(\frac{1}{T_B}-\frac{1}{T_A}\right),
\label{dots}
\end{equation}

where $j$ is the current. Figure~\ref{Entropy} represents  $\sigma$ as a function of the currents $j$ and it is colored according to $T_A$ (upper panels) and $T_B$ (lower panels). We analyze first the non coarse grained dynamics (on the left panels) where maxima of $\sigma$ are observed for each curve of $T_B$. Moreover, each curve has the maximum located at almost the same temperature $T_A$ which roughly corresponds to the critical temperature of the equilibrium Ising model $T_c$. Moreover, the maxima always fall in between the \textit{conductive} and \textit{convective} regimes. This is even more evident when analyzing the plots for the coarse grained dynamics (right panels) as the separation between these two regimes corresponds to a sharp change of $\sigma$ especially at lower temperatures $T_B$. In this case, the location of the maximum also corresponds to the critical temperature $T_c$.

This latter fact is indeed relevant for two reasons: first, it points to the fact that equilibrium properties and relevant quantities of equilibrium dynamics, e.g the critical temperature $T_c$,  still play an important role when the system is set out of equilibrium. Second it correlates to the phenomenon of self-organized criticality widely invoked to explain the fact that many physical systems set into non equilibrium states close to their critical temperature actually display ``exact'' critical properties~\cite{bak1988self,bak1990self}. Here the states obtained following a  maximum entropy production principle coincide with the ones obtained invoking the self organized criticality. 

\begin{figure}
\includegraphics[width=0.5\textwidth]{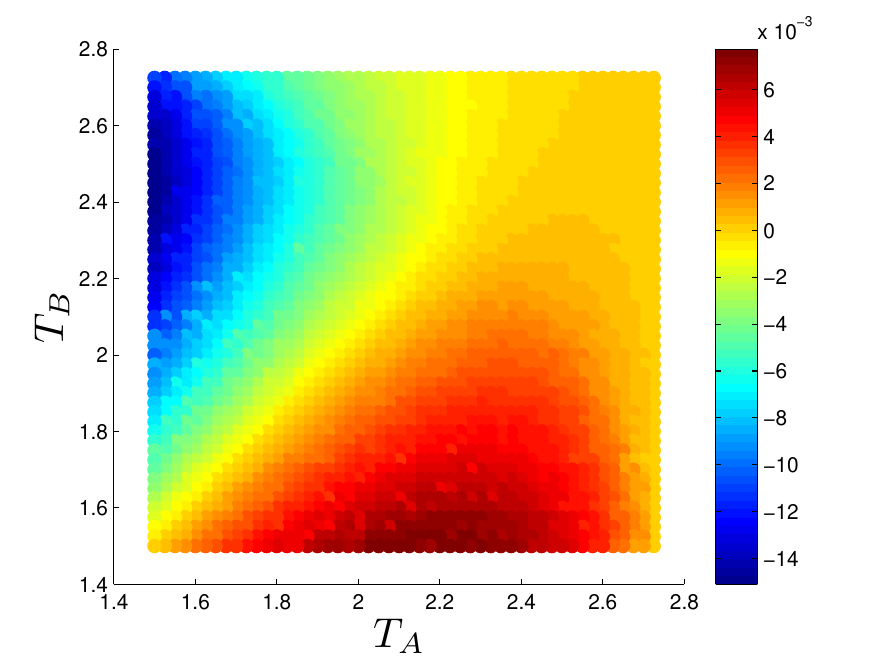}\includegraphics[width=0.5\textwidth]{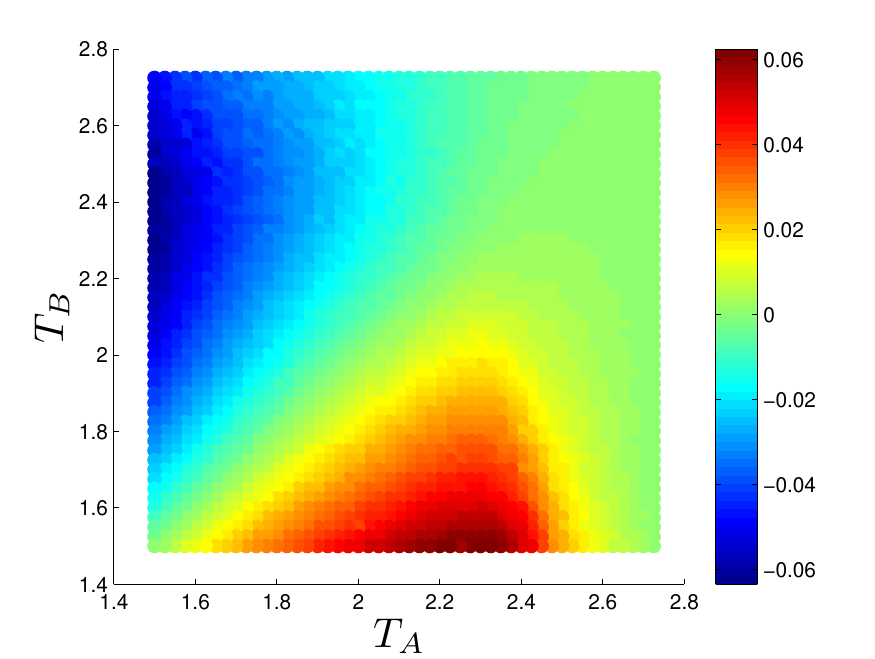}
\caption{Entropy production $\sigma$ (in color scale) as a function of the temperatures $T_A$  and $T_B$. Left panel:  lattice without coarse grain $L_A=L_B=64$. Right panel: coarse grain with $L_A=64, L_B=8$.} 
\label{EntropySpace}
\end{figure}

Finally, we can study the distribution of $\sigma$ in the plane $(T_A, T_B)$ as shown in Fig. \ref{EntropySpace}.  Note that the scales of $\sigma$ for the non coarse grained dynamics (left panel) is different from the one of the coarse grain simulations (right panels). The latter case shows a sharper $\sigma$ gradient than the former case. This is consistent with what is observed in Fig.~\ref{Entropy}.

The previous analysis suggests that the coarse graining step induces discontinuous changes in the system thermodynamic properties. We study this effect introducing a cascade of coarse grain operations and build a model of interacting lattices of different sizes (or, conversely, scales). In the numerical experiment, we start from a full resolution $L_A=64$ and halve it several times. At each scale we perform a dynamic Ising step. The intermediate resolutions considered are $L_B=32$, $L_C=16$ and $L_D=8$. We update lattice $A$ after the dynamical steps performed on $D$.\\

\begin{figure}
\includegraphics[width=0.75\textwidth]{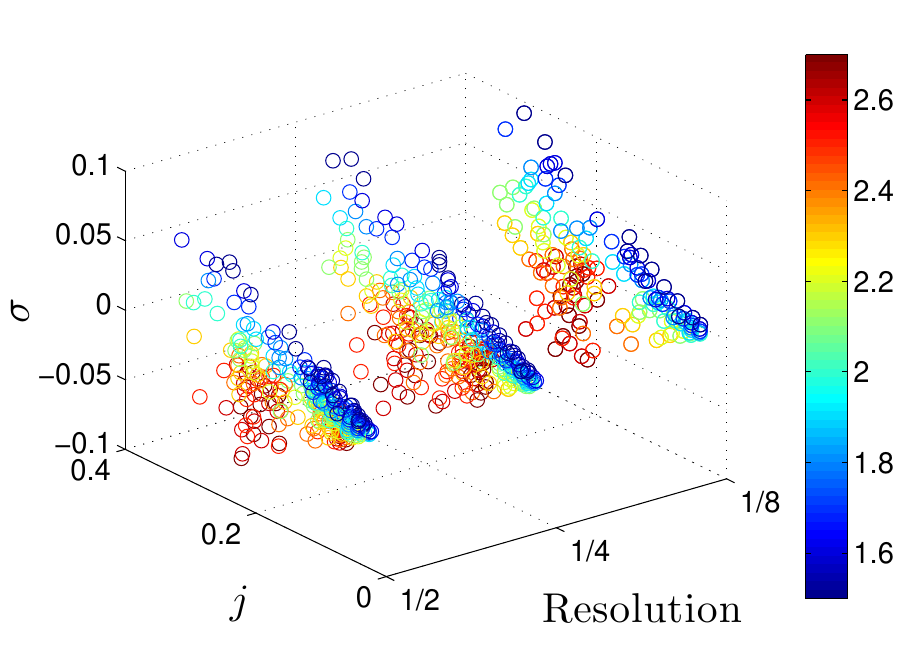} 
\caption{Entropy production $\sigma$ as a function of the currents and the change in the resolution.. The color scale depicts the temperature of the lattice at the lower resolution.} 
\label{entrscal}
\end{figure}

We first verify that the entropy production is constant through such a scale space. In order to do so we compare several values of $\sigma$ obtained at different change in resolutions. For example, we can compare   $\sigma_{A\to B}$ with $\sigma_{B\to C}$ and $\sigma_{C\to D}$ where the resolution has changed by half,  $\sigma_{A\to C}$ with $\sigma_{B\to D}$ and compare all of them with $\sigma_{A\to D}$ where the resolution changed by a factor 8. 
Results are shown in Fig.~\ref{entrscal}. One can observe that there is no substantial difference among the entropy productions defined.  This property is therefore reminiscent of the constant flux of energy observed in turbulence and necessary to explain the famous $k^{-5/3}$ energy spectrum.
However if one looks at the shape of such curves and compare them with the ones shown in Fig. \ref{Entropy}  several differences emerge. In the previous analysis we have performed an abrupt coarse graining shrinking the resolution by a factor 8 in one go. Here we have gone through several operations where the resolution has been halved. As a result of this smoother procedure, the curves of entropy production shown in figure~\ref{entrscal} show neither a clear maximum nor a transition between convective and conductive regimes. The convective regime is therefore closely linked to an abrupt change of lattice resolution (or scale).

\section{Discussion}
\label{disc}

In this paper we have obtained an effective way to introduce a genuine non equilibrium Ising dynamics with an interaction in the scale space reproducing some properties of turbulent systems.

Previously, other authors have investigated the possibility of building such a model by using only a one lattice dynamics and introducing the temperature gradient within the site of the lattice. Although such a system possesses a rich dynamics with an interesting variety of metastable and stable states, they do not allow for interaction among different scales and presents discontinuities in the temperature field which hinder simple tests of Onsager relations as well as out of equilibrium conjectures like the existence of maxima of entropy production. 

Our model satisfies such constraints by composing the dynamics of two interacting lattices at different temperatures. This model not only respects Fourier's law $dE/dS|_N=T_A$, but also features the possibility of having the second lattice at a coarse grained resolution, as it happens in several physical phenomena. Moreover, clear differences arise when comparing the coarse grained dynamics with the one of two lattices having the same spatial resolutions. For example, the coarse  grained dynamics exhibit a breakdown of Fourier's law (and therefore of conductive heat transfer) introducing, at high temperature gradients, convective fluxes confirmed by the appearance of a negative temperature in the energy-entropy relation. Moreover, the states maximizing the entropy production are located between conductive and convective regimes and correspond to the critical temperature of the equilibrium Ising dynamics. 

The convective states are new features for the Ising dynamics and cannot be observed e.g. by adding an external magnetic field to the system. In addition, the introduction of the temperature gradient between the two different lattices preserves the symmetry spin up/down which is broken with the addition of magnetic fields.

Despite its relative simplicity, our model displays coexistence of quasi-equilibrium states at the largest scale, with non-zero average fluxes at smaller scale. It therefore shows that turbulent flows may not be so exceptional, and rather be representative of a wide class of out-of-equilibrium systems with both spatial and scale dynamics.

An interesting outcome of our model could be the development of more realistic models of natural phenomena. For example, several models of atmospheric convection are based on the equilibrium Ising dynamics~\cite{majda2002stochastic,khouider2006simple}. We believe that a better representation of atmospheric transport can be achieved introducing our non equilibrium dynamics modeling both conductive and convective heat transfer. On the other hand, one can imagine that the second lattice is just some coarse grained properties of the first one and that temperature $T_B$ maps some properties in the scale space.

Finally, the fact that the maxima of entropy production correspond to states at critical temperature for the lattice suggests that there is a link between the maximum entropy production principle and self-organized criticality. The latter is often invoked to explain why out of equilibrium systems settle into states close to their critical manifolds but without a solid physical underlying motivation. We believe that such a connection might represent a step in solving this issue, despite requiring further investigations.

\section{Acknowledgments}

The authors acknowledge discussion and exchanges with S. Thalabard and B. Saint Michel. DF has been partially supported by the ERC Grant A2C2 (No. 338965)

\bibliography{isingOE}

\begin{thebibliography}{10}

\bibitem{bak1990self}
Per Bak.
\newblock Self-organized criticality.
\newblock {\em Physica A: Statistical Mechanics and its Applications},
  163(1):403--409, 1990.

\bibitem{bak1988self}
Per Bak, Chao Tang, and Kurt Wiesenfeld.
\newblock Self-organized criticality.
\newblock {\em Physical review A}, 38(1):364, 1988.

\bibitem{bertin2009far}
Eric Bertin and Olivier Dauchot.
\newblock Far-from-equilibrium state in a weakly dissipative model.
\newblock {\em Physical review letters}, 102(16):160601, 2009.

\bibitem{bouchet2012statistical}
Freddy Bouchet and Antoine Venaille.
\newblock Statistical mechanics of two-dimensional and geophysical flows.
\newblock {\em Physics reports}, 515(5):227--295, 2012.

\bibitem{cheng1991long}
Z~Cheng, PL~Garrido, JL~Lebowitz, and JL~Vall{\'e}s.
\newblock Long-range correlations in stationary nonequilibrium systems with
  conservative anisotropic dynamics.
\newblock {\em EPL (Europhysics Letters)}, 14(6):507, 1991.

\bibitem{derrida2002large}
B~Derrida, JL~Lebowitz, and ER~Speer.
\newblock Large deviation of the density profile in the steady state of the
  open symmetric simple exclusion process.
\newblock {\em Journal of statistical physics}, 107(3-4):599--634, 2002.

\bibitem{garrido1990long}
Pedro~L Garrido, Joel~L Lebowitz, Christian Maes, and Herbert Spohn.
\newblock Long-range correlations for conservative dynamics.
\newblock {\em Physical Review A}, 42(4):1954, 1990.

\bibitem{garrido1987ising}
PL~Garrido and J~Marro.
\newblock Ising models with anisotropic interactions: Stationary nonequilibrium
  states with a nonuniform temperature profile.
\newblock {\em Physica A: Statistical Mechanics and its Applications},
  144(2):585--603, 1987.

\bibitem{glauber1963coherent}
RJ~Glauber.
\newblock Time-dependent statistics of the ising model.
\newblock {\em Journal of Mathematical Physics}, 4(2):294--307, 1963.

\bibitem{huang2011entropy}
XL~Huang, B~Cui, and XX~Yi.
\newblock Entropy and specific heat for open systems in steady states.
\newblock {\em Modern Physics Letters B}, 25(03):175--183, 2011.

\bibitem{ising1925beitrag}
Ernst Ising.
\newblock Beitrag zur theorie des ferromagnetismus.
\newblock {\em Zeitschrift f{\"u}r Physik A Hadrons and Nuclei},
  31(1):253--258, 1925.

\bibitem{khouider2006simple}
Boualem Khouider and Andrew~J Majda.
\newblock A simple multicloud parameterization for convectively coupled
  tropical waves. part i: Linear analysis.
\newblock {\em Journal of the atmospheric sciences}, 63(4):1308--1323, 2006.

\bibitem{lebon2008understanding}
Georgy Lebon, David Jou, and Jos{\'e} Casas-V{\'a}zquez.
\newblock {\em Understanding non-equilibrium thermodynamics: foundations,
  applications, frontiers}.
\newblock Springer, 2008.

\bibitem{maes1991anisotropic}
Christian Maes and Frank Redig.
\newblock Anisotropic perturbations of the simple symmetric exclusion process:
  long range correlations.
\newblock {\em Journal de Physique I}, 1(5):669--684, 1991.

\bibitem{majda2002stochastic}
Andrew~J Majda and Boualem Khouider.
\newblock Stochastic and mesoscopic models for tropical convection.
\newblock {\em Proceedings of the National Academy of Sciences},
  99(3):1123--1128, 2002.

\bibitem{mccoy1973two}
Barry~M McCoy and Tai~Tsun Wu.
\newblock $\{$The Two-Dimensional Ising Model$\}$.
\newblock 1973.

\bibitem{mihelich2014maximum}
Martin Mihelich, B{\'e}reng{\`e}re Dubrulle, Didier Paillard, and Corentin
  Herbert.
\newblock Maximum entropy production vs. kolmogorov-sinai entropy in a
  constrained asep model.
\newblock {\em Entropy}, 16(2):1037--1046, 2014.

\bibitem{mihelich2014towards}
Martin Mihelich, Berengere Dubrulle, Didier Paillard, Corentin Herbert, and
  Davide Faranda.
\newblock Towards an understanding of the maximum entropy production principle
  in climate toy models.
\newblock In {\em EGU General Assembly Conference Abstracts}, volume~16, page
  15060, 2014.

\bibitem{onsager1949nuovo}
L~Onsager.
\newblock Nuovo cimento, 6.
\newblock {\em Suppl}, 2:249, 1949.

\bibitem{onsager1944crystal}
Lars Onsager.
\newblock Crystal statistics. i. a two-dimensional model with an order-disorder
  transition.
\newblock {\em Physical Review}, 65(3-4):117, 1944.

\bibitem{pinto2014critical}
Oscar~A Pinto, Federico Rom{\'a}, and Sebastian Bustingorry.
\newblock Critical behavior and out-of-equilibrium dynamics of a
  two-dimensional ising model with dynamic couplings.
\newblock {\em The European Physical Journal B}, 87(12):1--10, 2014.

\bibitem{pleimling2010convection}
Michel Pleimling, B~Schmittmann, and RKP Zia.
\newblock Convection cells induced by spontaneous symmetry breaking.
\newblock {\em EPL (Europhysics Letters)}, 89(5):50001, 2010.

\bibitem{praestgaard2000lattice}
Eigil~Luxh{\o}j Pr{\ae}stgaard, B~Schmittmann, and RKP Zia.
\newblock A lattice gas coupled to two thermal reservoirs: Monte carlo and
  field theoretic studies.
\newblock {\em The European Physical Journal B-Condensed Matter and Complex
  Systems}, 18(4):675--695, 2000.

\bibitem{thalabardstatistical}
Simon Thalabard, Brice Saint-Michel, Eric Herbert, François Daviaud, and
  Bérengère Dubrulle.
\newblock A statistical mechanics framework for the large-scale structure of
  turbulent von kármán flows.
\newblock {\em New Journal of Physics}, 17(6):063006, 2015.

\end{thebibliography}
\bibliographystyle{plain}

\end{document}